\documentclass[10pt]{article}
\usepackage{amsmath,bm,graphicx,amssymb,amsthm}
\usepackage{amsmath}
\usepackage{amsfonts}
\usepackage{amssymb}
\usepackage{graphicx}
\usepackage[dvips]{color}
\begin{document}

\begin{center}
{\LARGE Vacuum thin shell solutions in five-dimensional}

\medskip

{\LARGE Lovelock gravity}

\medskip

\medskip

\medskip

{\large C. Garraffo}$^{1,2}${\large , G. Giribet}$^{3,4}${\large ,
E. Gravanis}$^{4}${\large , S. Willison}$^{4}$

\medskip

$^{1}$ Instituto de Astronom\'{\i}a y F\'{\i}sica del Espacio,
CONICET, Argentina.

{\it Ciudad Universitaria, IAFE, C.C. 67, Suc. 28, 1428, Buenos
Aires, Argentina}.

\medskip

$^{2}$ Brandeis Theory Group, Martin Fisher School of Physics,
Brandeis University, 

{\it Brandeis University, Waltham, MA 02454-9110, USA}.

\medskip

$^{3}$ Departamento de F\'{\i}sica, FCEN, Universidad de Buenos
Aires, Argentina,

{\it Ciudad Universitaria, Pabell\'on 1, 1428, Buenos Aires,
Argentina}.

\medskip

$^{4}$ Centro de Estudios Cient\'{\i}ficos CECS,

{\it Casilla 1469, Valdivia, Chile.}

\medskip

\end{center}

\begin{abstract}
{Junction conditions for vacuum solutions in five-dimensional
Einstein-Gauss-Bonnet gravity are studied. We focus on those cases
where two spherically symmetric regions of space-time are joined in
such a way that the induced stress tensor on the junction surface
vanishes. So a spherical vacuum shell, containing no matter, arises
as a boundary between two regions of the space-time. Such solutions
are a generalized kind of spherically symmetric empty space
solutions, described by metric functions of the class $C^0$. New
global structures arise with surprising features. In particular, we
show that vacuum spherically symmetric wormholes do exist in this
theory. These can be regarded as gravitational solitons, which
connect two asymptotically (Anti) de-Sitter spaces with different
masses and/or different effective cosmological constants. We prove
the existence of both static and dynamical solutions and discuss
their (in)stability under perturbations that preserve the symmetry.
This leads us to discuss a new type of instability that arises in
five-dimensional Lovelock theory of gravity for certain values of
the coupling of the Gauss-Bonnet term. }
\end{abstract}

\newpage

A higher dimensional theory which has attracted much interest is
Lovelock gravity \cite{Lovelock}. This is because the theory, having
field equations of second order in derivatives of the metric,
intuitively has the right ingredients for a classical theory of
gravity. In particular, the linearised perturbations about physically sensible backgrounds
are well-behaved and are of the same second derivative form as in General Relativity (GR).
Equivalently, the quadratic part of the perturbed Lagrangian is of
the general form $h \partial
\partial h$ so there are no corrections to the propagator and no
extra (ghost) fields corresponding to higher
derivatives \cite{Zwiebach,Zumino}.

There are however some exotic features of Lovelock gravity which
certainly do not arise in GR. One such feature is the problem of
(non-)determinism \cite{Teitelboim-87,Choquet-Bruhat-88,Deruelle-03}. Given an
initial data surface and a specified intrinsic metric and its first
time derivative (or extrinsic curvature) one can try to integrate
the Lovelock equations to evolve the metric through time. There one
runs into a theoretical problem: There are solutions with spacelike
surfaces on which the extrinsic curvature may be suddenly
discontinous. It can not be determined from the initial data if the
extrinsic curvature will jump or if the metric will evolve smoothly.
This is equivalent to the nonuniqueness problem in inverting the
canonical momentum which is polynomial in the
curvature \cite{Teitelboim-87}. Even for more smooth metrics there
can be a problem of indeterminism, where components of the metric
become arbitrary. This second kind of nondeterminism, with arbitrary
functions of time appearing, only occurs in a regime where the
curvature is large enough that the higher order Lovelock tensors
become appreciable compared to the Einstein tensor. However the
first kind of indeterminism, for metrics of class $C^0$, is quite
generic in Lovelock gravity. This means
that one has to be careful in interpreting Lovelock theory as an
effective theory. It is too simplistic to say that the theory is
valid when the curvature is small w.r.t. a certain characteristic
scale.

A natural question arises: can we look at the same phenomenon in the
context of timelike surfaces. That is, discontinuities allowed in
integrating the equations of motion in a spacelike direction. The
nonsmooth solutions we shall present here (first found in Ref.
\cite{GGGW}) are the timelike analogues of the first kind of
nondeterminism. These objects are not a priori pathological objects
in the theory: they can be everywhere non-spacelike (they can even
be static as we shall see) and so in principle they do not violate
determinism. One of the original motivations for this work was to
see whether stable solitonic objects can exist in a space which at
large distances looks like a positive mass solution of GR (such
objects might be interpreted as branes of the Lovelock theory
itself). It seems that the answer is no and the reasons why they do
not exist are interesting in their own right.
\\

The essential features can be seen in the quadratic Lovelock theory,
 often called Einstein-Gauss-Bonnet (EGB) theory. We shall therefore
 restrict ourselves to this theory and to the minimum number of dimensions, i.e. five. The
 action is given by the Einstein-Hilbert term,
plus the Einstein cosmological term and additionally the
Gauss-Bonnet combination of quadratic curvature invariants:

\begin{equation}
S=\frac{1}{2\kappa ^{2}}\int d^{5}x\sqrt{-g}\left( {\mathcal R}-2\Lambda
+\alpha \left( {\mathcal R}^{2}+{\mathcal R}_{A B C D }{\mathcal R}^{A B C D
}-4{\mathcal R}_{A B }{\mathcal R}^{A B }\right) \right) ,\label{The_Action}
\end{equation}%
where $\kappa ^{2}=8\pi G$ and $\alpha $ represents the coupling
constant of the Gauss-Bonnet term. In five dimensions, this is in
fact the most general Lovelock theory since the Lovelock combination
of cubic terms $\sim O({\mathcal R}^{3})$ identically vanishes (in $D=6$ they
combine to a quantity which is locally a total derivative; they
contribute to the equations of motion for $D\geq7$; see for instance \cite{Garraffo2}). Likewise, the
$n$th order Lovelock terms only become relevant in $2n+1$ or more
dimensions.

The field equations associated with the action (\ref{The_Action})
coupled to some matter action take the form
\begin{gather}\label{The_field_equations}
 G^A_{B} + \Lambda \delta^A_B + \alpha H^A_B
 = \kappa^2 T^A_B\,,
\end{gather}
where $T^A_B$ is the stress tensor, $G^A_B  \equiv -\frac{1}{4}\,
\delta^{A CD}
 _{B EF}\, {\cal R}^{EF}_{AB}
 = {\cal R}^A_{\ B}- \frac{1}{2} \delta^A_B\, {\cal R}$ is the
 Einstein tensor, and
\begin{align*}
 H^A_B & \equiv -\frac{1}{8}\, \delta^{A C_1 \dots C_4}
 _{B D_1 \dots D_4}\, {\cal R}^{D_1 D_2}_{\ \ \ C_1
 C_2} {\cal R}^{D_3 D_4}_{\ \ \ C_3 C_4} \, ,
\end{align*}
and where the antisymmetrized Kronecker delta is defined as
$\delta^{A_1 \dots A_p}_{B_1 \dots B_p} \equiv p!
\delta^{A_1}_{[B_1} \cdots  \delta^{ A_p}_{B_p]}$.

The spherically symmetric solution in this theory with $T_{AB} =0$,
i.e. the analog to the Schwarzschild
black hole in Einstein's Theory, is the Boulware-Deser solution, which reads 
\cite{Boulware:1985wk,Wheeler,Cai:2001dz}
\begin{equation}\label{BD_metric}
ds^{2}=-f(r)dt^{2}+\frac{1}{f(r)}dr^{2}+r^{2}d\Omega _{3}^{2}, \ \ f(r)=k+\frac{r^{2}}{4\alpha}\left(  
1+\xi\sqrt{1+\frac
{4\Lambda\alpha}{3}+\frac{16 M \alpha}{r^{4}}}\right)
\end{equation}%
where $ d\Omega _{3}^{2}=\sin ^{2}\chi d\theta ^{2}+\sin ^{2}\chi
\sin ^{2}\theta d\phi ^{2}+d\chi ^{2} $ is the line element of the
three-sphere with normalized curvature $k= 1$ (solutions also exist with
planar and hyperbolic horizon geometry, i.e. with $k=0$, $-1$, respectively. For simplicity, we
will focus here on the spherical case $k=1$) and
 $\xi^2=1$.

We see here a typical feature of the EGB theory: the Boulware-Deser
\cite{Boulware:1985wk} metric has two \emph{branches}. The minus branch ($\xi=-1 $) reduces
to the corresponding solution of GR in the limit
$\alpha\rightarrow0$, as expected. However, for the plus branch
($\xi=+1$) this limit is ill defined. Thus, the plus branch is
called the ``exotic branch'' of the Boulware-Deser metrics and it is
usually thought of as an unstable vacuum of the theory, with ghost
excitations \cite{Boulware:1985wk,Zwiebach}, and a naked singularity
instead of a black hole. Just as for Schwarzchild's metric, $M$ is
here a constant of integration and it is associated with the mass of
the solution. Let us also point out that the Boulware-Deser solution
is unique only under a certain assumption about the coupling
constants (in the case of 5-dimensional EGB theory the assumption is
$4\alpha/3\Lambda \neq 1$) discussed in\footnote{See also \cite{Dotti},
were the non-uniqueness of the solution at the point of the space parameters
$\Lambda \alpha =-3/4$ is analysed.} Refs. \cite{Wheeler,
Charmousis,Zegers,DF} and also the assumption that the metric is of class
$C^2$ \cite{Zegers}. It is the relaxation of this last assumption
which we explore in this article.

The spherically symmetric situation gives a simple setting in which
to construct some intriguing vacuum geometries which are special to
Lovelock gravity: we can construct thin-shell vacuum wormholes and
other objects by gluing together different Boulware-Deser metrics.
In order to study these geometries we will start by discussing the
junction conditions in this theory, worked out in
\cite{Davis:2002gn,Gravanis:2002wy}. These are the analogues of the
Israel conditions in GR \cite{I}. In particular,
they will be employed to join two different spherically symmetric
spaces.

Let $\Sigma$ be a timelike hypersurface separating two bulk regions
of spacetime, region ${\cal V}_L$ and region ${\cal V}_R$ (``left"
and ``right"). We introduce, for convenience, the coordinates
($t_L,r_L$) and ($t_R,r_R$) and the metrics
\begin{equation}
ds_{L}^{2}=-f_{L}\,dt_{L}^{2}+\frac{dr_{L}^{2}}{f_{L}}+r_{L}^{2}d\Omega^{2}_{3}\, ,\label{bulk L}%
\end{equation}
\begin{equation}
ds_{R}^{2}=-f_{R}\,
dt_{R}^{2}+\frac{dr_{R}^{2}}{f_{R}}+r_{R}^{2}d\Omega^{2}_{3} \label{bulkR}\,
,\end{equation} in the respective regions. We are interested in the
case where both $f_L(r_L)$ and $f_R(r_R)$ are vacuum solutions, so
they will be of the form given in equation (\ref{BD_metric}). In
general, the mass parameter $M_R$ will be different from $M_L$, and
$\xi _R$ different from $\xi _L$ so that the two different branches
of the Boulware-Deser solution can be joined.

It is also convenient to parameterize the shell's motion in the
$r-t$ plane using the proper time $\tau$ on $\Sigma$. In region
${\cal V}_L$ we have $r_{L}=a(\tau)$, $t_{L}=T_{L}(\tau)$ and in
region ${\cal V}_R$ we have $r_{R}=a(\tau)$, $t_{R}=T_{R}(\tau)$.
The induced metric on $\Sigma$ induced from region ${\cal V}_L$ is
the same as
that induced from region ${\cal V}_R$, and is given by%
\begin{equation}
d\hat{s}^2=-d\tau^{2}+a(\tau)^{2}d\Omega^{2}_{3}\, .\label{4-geometry}%
\end{equation}
 This guarantees the existence of a coordinate system where the
metric is continuous ($C^0$). Let us set some conventions: The
hypersurface $\Sigma$ has a single unit normal vector $\bm{n}$ which
points from left to right; and the orientation factor $\eta$ of each
bulk region is defined as follows: $\eta = +1$ if the radial
coordinate $r$ points from left to right, while $\eta = -1$ if the
radial coordinate $r$ points from right to left.

We are now in position to classify the shells according to the
following definitions: $\eta_L\eta_R>0$ will be called the standard
orientation; $\eta_L\eta_R<0$ will be called the wormhole
orientation\footnote{Notice that this geometry could correspond to
joining two ``exterior regions'' of a spherical solution as well as
two ``interior regions''.}.

Integrating the field equations from left to right in an
infinitesimally thin region across $\Sigma$ one obtains the junction
conditions. This relates the discontinuous change of spacetime
geometry across $\Sigma$ with the stress tensor $S_{a}^{b}$ (see
Refs. \cite{Davis:2002gn,Gravanis:2002wy,Gravanis:2007ei} for
details).
\begin{equation}
\label{explicit junction ij}   ({\frak Q}_R)_{a}^{b}-({\frak
Q}_L)_{a}^{b} =- \kappa^{2} S_{a}^{b}\ ,
\end{equation}
Above, the subscripts $L$, $R$ signify the quantity evaluated on
$\Sigma$ induced by regions ${\cal V}_L$ and ${\cal V}_R$
respectively. The symmetric tensor ${\frak Q}^a_{\ b}$ is given by
\begin{gather}\label{explicit_Q}
 {\frak Q}^a_{\ b} =  - \delta^{ac}_{bd} K^d_c +\alpha\, \delta^{acde}_{bfgh}
 \Big( -  K^f_c R^{gh}_{\ \ de}
 + \frac{2}{3} K^f_cK_d^gK_e^h \Big)\, ,
\end{gather}
where $a$, $b$,... are indices on the tangent space of the
world-volume of the shell. The symbol $K^a_b$ refers to the
extrinsic curvature, while the symbol $R^{ab}_{\ \ cd}$ appearing
here corresponds to the four-dimensional intrinsic curvature (see
\cite{GGGW} for details). Once applied to the spherically symmetric
case the tensor ${\frak Q}_{a}^{b}$ turns out to be diagonal with
components
\begin{align}
{\frak Q}^{\tau}_{\tau} & = -3  \ a^{-3}   \bigg( \eta\ a^{2}\,
\sqrt{\dot a^{2}+f} +\, 4 \alpha\, \eta\,\sqrt{\dot a^{2}+f} \
\big(k+\frac{2}{3} \dot a^{2}-\frac
{1}{3}f\big) \bigg) \ ,\label{Qtt}\\
{\frak Q}_\theta^\theta & = {\frak Q}_\chi^\chi = {\frak
Q}_\varphi^\varphi\, .
\end{align}

It can be verify that the following equation is satisfied
\begin{equation}
\label{Q conservation}\frac{d}{d\tau}\big(a^{3} {\frak Q}^{
\tau}_{\tau}\big)=\dot a \, 3 a^{2} {\frak Q}_{\theta}^{\theta}\ ,.
\end{equation}
This equation expresses the conservation of $S_{a}^{b}$, i.e. no
energy flow to the bulk, which always holds when the
normal-tangential components of the energy tensor in the bulk is the
same in both sides of the junction
hypersurface \cite{Davis:2002gn,Gravanis:2007ei}.

The main point here is that non-trivial solutions to (\ref{explicit
junction ij}) are possible even when $S_{a}^{b}=0$. That is, the
extrinsic curvature can be discontinuous across $\Sigma$ with no
matter on the shell to serve as a source. The discontinuity is then
self-supported gravitationally and this is due to non-trivial
cancelations between the terms of the junction conditions. Similar
configurations are impossible in Einstein gravity (in that case the
junction conditions are linear in the extrinsic curvature). Since we
are interested in vacuum solutions, we will consider
\begin{gather}
S_{a}^{b}=0\ .
\end{gather}

\noindent From equation (\ref{Q conservation}) we see that in the
case $\dot a \ne0$, the components of the junction condition are not
independent: $ ({\frak Q}_{R})_{\tau}^{\tau}-({\frak
Q}_{L})_{\tau}^{\tau}=0$ $\Rightarrow$ $({\frak
Q}_{R})_{\theta}^{\theta}-({\frak Q}_{L})_{\theta}^{\theta}=0 \ . $
So it suffices to impose only the first condition, which can be
factorized as follows,
\begin{align}
&\left(\eta_R \sqrt{\dot{a}^{2}+  f_R } - \eta_L \sqrt{\dot{a}^{2}+
f_L }\right)\times    \nonumber
\\&\qquad \times\left\{ a^{2} +
4\alpha(k+ \dot{a}^{2}) - \frac{4\alpha}{3} \left( f_{R} + f_{L} + 2
\dot
{a}^{2} + \eta_{R} \eta_{L} \sqrt{ f_{R} +\dot{a}^{2}} \sqrt{ f_{L} +\dot{a}%
^{2}}\right) \right\}  =0\, .    \label{explicit general junction}
\end{align}

\noindent All the information concerning the spherically symmetric
vacuum shells is contained in (\ref{explicit general junction}).
There exist several possibilities to be explored, corresponding to
different choices in the Bolware-Deser parameters $k,\ M$ and $\xi
$, combined with the two possible orientations $\eta $. This
permits a very interesting catalogue of geometries which we survey
later and is further explored in \cite{GGGW}.

The first factor in (\ref{explicit general junction}) vanishes
for the smooth metric. Thus, for non-smooth solutions we demand that
the second factor vanishes. From the second factor, squaring
appropriately, we obtain
\begin{equation}
\label{solved-explicit} \dot a^{2}=\ \sigma \,
\frac{\Big(f_{R}+f_{L} -3(k+a^{2}/4\alpha) \Big)^{2}-f_{R}f_{L}} {3
\Big(f_{R}+f_{L}-2(k+a^{2}/4\alpha) \Big)}=:- V(a) \, ,
\end{equation}
This is essentially a one-dimensional problem, given by an ordinary
differential equation (\ref{solved-explicit}), like the equation for
a particle of a given energy moving radially in a spherical potential.
Now, since we have squared the junction condition, we must
substitute (\ref{solved-explicit}) back into (\ref{explicit general
junction}) to check the consistency. When doing so we find the
following restrictions
\begin{equation}
-\eta _{R}\eta _{L}\ (2f_{R}+f_{L}-3(k+a^{2}/4\alpha ))\
(2f_{L}+f_{R}-3(k+a^{2}/4\alpha ))\geq 0\, ;  \label{timed ineq}
\end{equation}
\begin{equation}
(f_{R} + f_{L} -2(k+a^{2}/4\alpha))>0\ . \label{timelike_real_roots}
\end{equation}
So, for a dynamical vacuum shell with a timelike world-volume
$\Sigma$, the scale factor of the metric (\ref{4-geometry}) on
$\Sigma$ is governed by (\ref{solved-explicit}), under the
inequalities (\ref{timed ineq}) and (\ref{timelike_real_roots}).

Using the inequalities we immediately obtain the following \emph{general results for dynamical or static shells:}
\\\\
G1)
 Vacuum shells with the standard orientation always
 involve the gluing of a plus branch $(\xi = +1)$ metric with a
 minus branch $(\xi = -1)$ metric.
\\\\
G2) Vacuum shells which involve the gluing of two
 minus branch $(\xi = -1)$ metrics exist only
  when the Gauss-Bonnet coupling constant $\alpha$
 satisfies $\alpha<0$.
 They always have the wormhole orientation.
\\

In the analysis above it has been explicitly assumed that $\dot{a}
\neq 0$.
It can be checked that, as expected, all the information
about the constant $a$ solutions can be obtained from the dynamical
case by imposing both $V(a_0)=0$ and $ V'(a_0)=0$.
Nevertheless, since the case $\dot{a}=0$ describing static
shells is of considerable interest, we shall treat it here explicitly.

So, let us now discuss the solutions for constant $a$, $a = a_0$. The bulk
metric in each of the two region is assumed to be of the
Boulware-Deser form (\ref{BD_metric}). In
this case the shell is located at fixed radius $r_L= r_R=a_0$. The
proper time on the shell's world-volume is $\tau = t_L\sqrt{f_L(a)}=
t_R\sqrt{f_R(a)}$ so that the induced metric on $\Sigma$ turns out
to be $d\hat{s}^2 = -d\tau^2 + a_0^2 d\Omega^2_3$. Then, the extrinsic
curvature components are $ K_{\tau}^{\tau}   = \eta\frac{f^{\prime}}
{2\sqrt{f}}$, $K_{\theta}^{\theta}
=K_{\chi}^{\chi}=K_{\varphi}^{\varphi}= \frac{\eta \sqrt{f}}{a}$ and
the intrinsic curvature components are $R_{\ \
\theta\varphi}^{\theta\varphi} =k/a_0^{2}$, etc. The junction conditions with $S^a_{b} = 0$ give:
\begin{align}
\label{static j 00}&S^\tau_\tau =0\ \Rightarrow &\big( \eta_{R}
\sqrt{f_{R}}- \eta_{L} \sqrt{f_{L}} \big)
\Big(a_{0}^{2}+\frac{4\alpha}{3} \big\{3k- f_{R} - f_{L} - \eta_{L}
\eta_{R} \sqrt{f_{L} f_{R}} \big\} \Big)=0\ ,
\\
\label{static j ij} &S^\theta_\theta = 0\ \Rightarrow \ &
\Big(\frac{\eta_{R}}{\sqrt{f_{R}}}-\frac{\eta_{L}}%
{\sqrt{f_{L}}}\Big) \Big(k-\frac{\Lambda a_{0}^{2}}{3}-\eta_{L}
\eta_{R} \sqrt{f_{L} f_{R}} \Big)=0\ ,
\end{align}
In both equations (\ref{static j 00}) and (\ref{static j ij}), the
first factor vanishes if and only if the metric is smooth. Again,
rejecting this as the trivial solution, we demand that the second
factor vanishes in both equations
(under the condition $f_L, f_R>0$).

Let us first consider $\Lambda \neq 0$.
Solving the equations we see that $f_L$ and
$f_R$ obey the same quadratic equation where one $f$ has the $+$
root of the solution and the other has the $-$ root. We will call
these solutions $ f_{+}$ and $ f_{-}$ respectively with corresponding parameters
$\xi_+$, $\xi_-$ and $M_+$, $M_-$.
Substituting the explicit expression for $f_{L,R}$, evaluated at $r = a_0$, we find
\begin{gather}
1+x \pm \sqrt{3} \sqrt{x(1+x)\Big(\frac{3}{x}+\frac{12}{\Lambda a_0^2}-1\Big)} =2
\xi_{(\pm)}
\sqrt{1+x+\frac{9 x^2  M_{(\pm)}}{\alpha \Lambda^2 a_0^4}}\, ,\label{General_Solution_mess_1}%
\end{gather}
where we have found it convenient
to define the dimensionless parameter\footnote{This parameter is important in determining the nature of the Boulware-Deser solutions.
For $x<-1$ both branches are pathological, with branch singularities where the metric becomes complex. In particular
there is no asymptotic region since the metric always becomes complex for $r \to \infty$. $x = -1$ is the special case,
related to the Chern-Simons theory of gravity in five dimensions, where the effective cosmological constants
of the two branches are the same. For $x> -1$ the $(-)$ branch solution is somewhat similar to the Schwartzschild/Schwartzchild-(A)dS
black hole. }
\begin{equation*}
x\equiv\frac{4\alpha\Lambda}{3} \, .
\end{equation*}

For a solution to exist, the square root in the l.h.s. of
(\ref{General_Solution_mess_1}) must be real, and since we have
squared the equations we must substitute back to check the
consistency. So we get (\ref{General_Solution_mess_1}) along with
the following inequalities:
\begin{equation}\label{causality condition}
 \frac{3}{x\Lambda a_0^2}(3+x)+2>0 \qquad
 \text{(Timelike shells)}\, ;
\end{equation}
\begin{equation}\label{orientation condition}
 \frac{3}{\Lambda a_0^2} <1 \quad \text{(Standard orientation)}\,, \qquad  \frac{3}{\Lambda a_0^2}  > 1 \qquad
 \text{(Wormhole orientation)}\, .
\end{equation}
These admit solutions for a wide range of the coupling constants $\Lambda$, $\alpha$ and parameters $\xi_\pm$,
$M_\pm$, which is described exhaustively in ref \cite{GGGW}. Here we mention some \emph{general results 
for static shells:} \\\\
S1) Static shells with wormhole orientation only exist for
$\Lambda >0$.
\\\\
S2) Static shells with wormhole orientation containing two
asymptotic regions only exist for $\alpha >0$. At least one region
will be asymptotically Anti-de Sitter.
\\\\
S3) Let $\Lambda \leq 0$. Then static shells exist (with
standard orientation) joining (+) with (-) branches.
\\

In S2) we used the fact that the metric is well defined as $r\to \infty$ only
for $1+4\alpha\Lambda/3>0$. Also, in S1-S3) we have preemptively written the result for $\Lambda = 0$  which we now show.
This is an interesting special case, in which the equations reduce to
\begin{align}
f_{L} + f_{R} & =2+\frac{3 a_{0}^{2}}{4 \alpha}\ ,\\
 \eta_L\eta_R\sqrt{f_{L} f_{R}} & =1\ .
\end{align}
We see from the second equation that $\eta_L \eta_R$ must be $+1$,
i.e. static wormholes do not exist for $\Lambda =0$. One can also
check that the consistency of the solutions leads to the condition
$\alpha >0$ as well as the $\Lambda=0$ case of S3).
\\

\begin{figure}[t]
 \begin{center}
  \includegraphics[angle=270,width=4.8in]{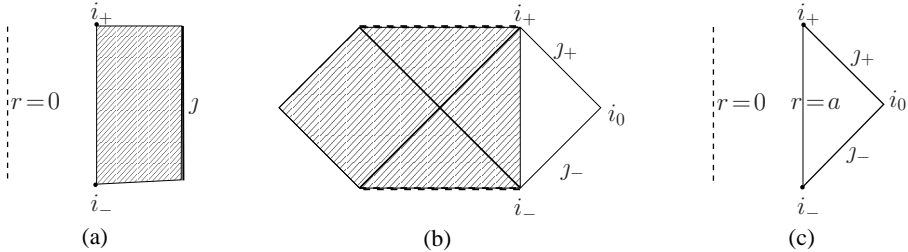}
  \caption{
An example of a solution with standard orientation, for
$\Lambda = 0$.
 a) (+) branch spacetime: naked singularity, asymptotically AdS;
  b) (-) branch spacetime: black hole, asymptotically flat;
  c) By cutting out shaded regions and joining we obtain a $C^0$ vacuum solution
  with a ``false vacuum bubble" containing a naked singularity (singularities are shown as dashed lines;
 the faint timelike line is the shell worldvolume).}
\label{badpic}
 \end{center}
\end{figure}

Summarizing, for the standard orientation geometries, $\eta _L \eta
_R>0$, branches are always $(\xi_{(-)} ,\xi _{(+)})=(-1,+1)$, so
region ${\cal V}_L$ has a different effective cosmological constant
to region ${\cal V}_R$, as can be seen from the expansion of the
metric for large $r$. In this sense the shell is like the false
vacuum bubbles studied in Refs. \cite{Vacuum bubbles}, but for a
false vacuum which is of purely gravitational origin (see
\cite{GGGW}). These kind of solutions might lead to curious
implications for the global spacetime structure. For instance, we
can construct a vacuum solution whose geometry, from the point of
view of an external observer, would coincide with that of a black
hole but, instead, would not possess a horizon. A particle in free
fall would not find a horizon but rather a naked singularity as soon
as it passes through the $C^0$ junction hypersurface located at
$r=a>r_H$. This is depicted in Fig. \ref{badpic} for the case $\Lambda =0$
(similar solutions also exist for $\Lambda \neq 0$). However, as one
would expect, such cosmic-censorship-spoiling shells are unstable
with respect to small perturbations, as we shall see below.

On the other hand, there are two different classes of wormhole
orientation geometries. The first class describes actual wormholes,
presenting two different asymptotic regions which are connected
through a throat located at radius $r_L = r_R =a$; the radius of the
throat being larger than the radius where the event horizons (or
naked singularities) would be. This type of geometry is an example
of a vacuum spherically symmetric wormhole solution in Lovelock
theory and its existence is a remarkable fact on its own. The second
class of wormhole-like geometry has no asymptotic regions, and is
obtained by cutting away the exterior region of both geometries and
gluing the two interior regions together.
\begin{figure}[t]
\begin{center}
  \includegraphics[width=.8\textwidth]{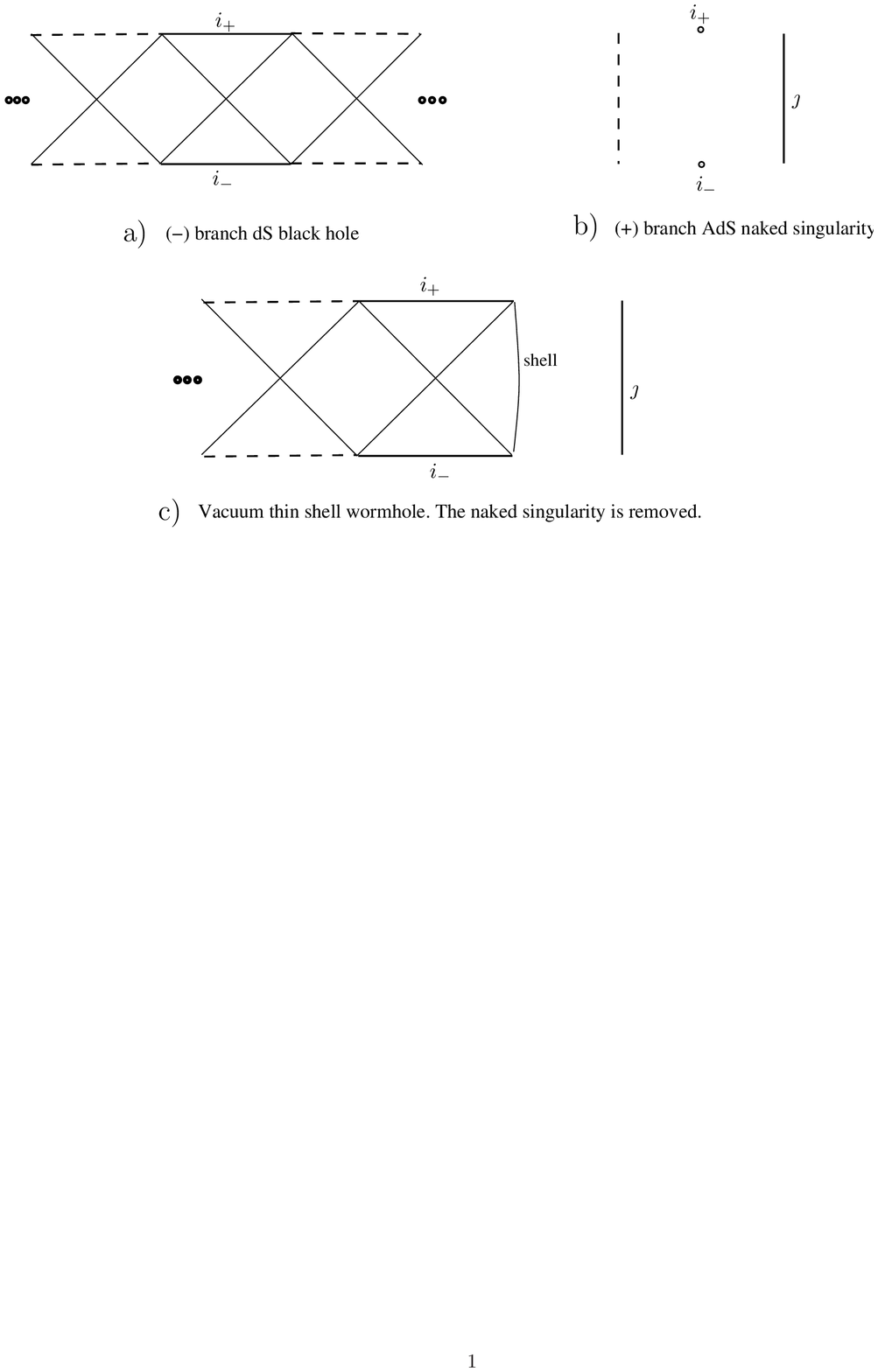}\\
\caption{ diagram c) shows a static wormhole joining two asymptotic dS/AdS
regions. This is a vacuum solution of Einstein-Gauss-Bonnet theory.}
   \end{center}
\end{figure}

\bigskip

Finally, we discuss dynamical shells and  the issue of stability of the static shells.
In general, vacuum shells will be dynamical objects. In order to discuss their dynamics
and stability let us briefly recapitulate upon the equation
(\ref{solved-explicit}), which governs the dynamics of the shells.
It takes the form:
\begin{equation}
 \dot a^{2}+ V(a) = 0 \, ;  \label{tonga2}
\end{equation}
(see (\ref{solved-explicit}) above). It is useful to introduce
the non-negative quantity $ Y= \sqrt{1+ \frac{4\alpha
\Lambda}{3}+ \frac{16M \alpha}{a^4}}$, with which the effective
potential reads
\begin{equation}
 V(a)=  \left(1+\frac{a^2}{4\alpha}\right)
 - \frac{ a^2}{4\alpha}
 \left(  \frac{3(\xi_R  Y_R + \xi_L Y_L )^2 + (\xi_R  Y_R - \xi_L Y_L )^2 }
 {12(\xi_R Y_R + \xi_L Y_L)}\right) \, . \label{tonga}
\end{equation}
In addition to the differential equation, the solution must obey the
inequalities (\ref{timed ineq}) and (\ref{timelike_real_roots}).

To analyze the motion of a shell we need to know the derivatives of
the potential (this is worked out in the appendix of \cite{GGGW}).
Differentiating the potential we get the following expression for
the acceleration of a moving shell,
\begin{equation}\label{force}
\ddot a=-\,\frac{a}{4\alpha} \Big[1-\frac{1 + 4\alpha\Lambda/3
}{\xi_R Y_R + \xi_L Y_L}\Big]\ .
\end{equation}
Considering the sign of this acceleration, we can make some general
observations: When $1 + \frac{4\alpha\Lambda}{3} \geq 0$ and $\alpha
<0$ a vacuum shell always experiences a repulsive force away from $r
= 0$; conversely when $1 + \frac{4\alpha\Lambda}{3} \leq 0$ and
$\alpha> 0$ a vacuum shell always experiences an attractive force
towards $r = 0$. Which means that if $\Sigma$ is a timelike shell it
will either be in an (unstable) static state, or, if it is moving,
will either expand or collapse, it can not be bound. 
\\

\begin{figure}[t]
 \begin{center}
  \includegraphics[width=2.2in]{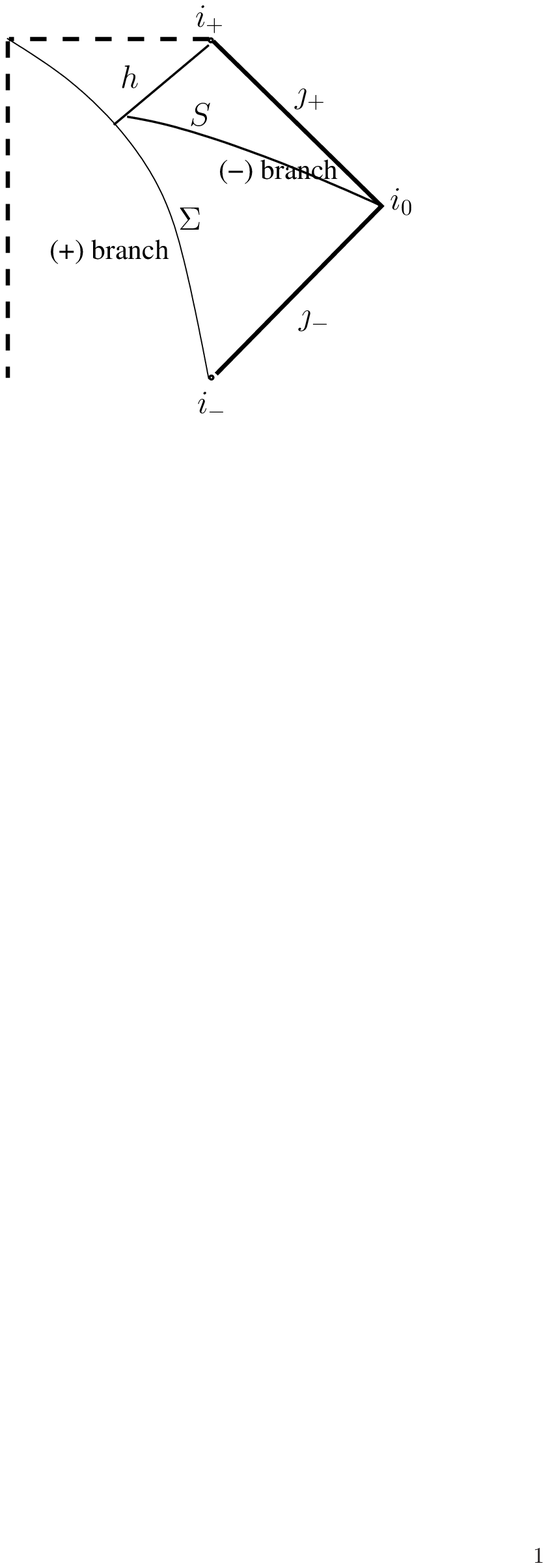}
 \end{center}
  \caption{ A typical example for $\Lambda =0$. Collapse of a vacuum shell with worldvolume represented by the line $\Sigma$.
  The (+) branch region shrinks inside the horizon $h$.
  ``Cosmic censorship" is restored in the future
 domain of dependence of the spacelike initial data surface $S$. }
\label{typical}
\end{figure}

Combining with
results derived from the inequalities we can state \emph{ further results for dynamical shells in the regime
$1+4\alpha\Lambda/3>0$:}
\\\\
D1) When $\alpha <0$ a vacuum shell always experiences a
repulsive force away from $r = 0$.
\\\\
D2) A shell joining two minus branches always experiences a
repulsion away from $r=0$.
\\\\
D3) A shell joining a minus branch with a plus branch region
will either: i) be in
an (unstable) static state or, after at most one bounce:
ii) collapse without reexpanding or
iii) expand indefinitely. It will not perform oscillations or
any other bounded motion.
\\\\
A corollory of D3) is that shells with standard orientation are
unstable (Fig. \ref{typical}).
\\

So in summary, we have found some general results for the range of
parameters $1+\frac{4\alpha\Lambda}{3} \geq 0$. This range is
of importance as it includes the case $|\alpha\Lambda| << 1$ and
therefore applies when the Gauss-Bonnet term is a small correction.
Combining these results, we conclude that, in this range of
parameters, all timelike vacuum shells involving the minus branch
are unstable. The only vacuum shell solutions which can be static or
oscillatory are wormholes which match two regions of the exotic plus
branch.

Here we have focused on the case where the shell is a 3-sphere
evolving though time in a spherically symmetric background. However
this analysis can be straightforwardly extended to the cases of any
constant curvature 3-manifold shell and to shells of either
spacelike or timelike signature. This generalization and a more
complete analysis of the space of solutions can be found in Ref.
\cite{GGGW}.

\[
\]

This paper is an extended version of the authors' contribution to the proceedings of the 12th Marcel Grossman Meeting, 
held in Paris, 12-18 July 2009. This work was supported in part by Fondecyt grant 1085323, by UBA grants UBACyT-X861 
UBACyT-X432, and by ANPCyT grant PICT-2007-00849. The work of C.G. is supported by CONICET. The Centro de Estudios 
Cient\'{\i}ficos CECS is funded by the Chilean Government through the Millennium Science Initiative and the Centers of 
Excellence Base Financing Program of Conicyt. CECS is also supported by a group of private companies which at present 
includes Antofagasta Minerals, Arauco, Empresas CMPC, Indura, Naviera Ultragas and Telef\'{o}nica del Sur.


\begin{thebibliography}{99}


\bibitem{Lovelock} D. Lovelock, J. Math. Phys. \textbf{12}, 498 (1971).


\bibitem{Zwiebach} B.~Zwiebach,
Phys.\ Lett.\ B \textbf{156} (1985) 315. 



\bibitem{Zumino} B.~Zumino,
Phys.\ Rept.\ \textbf{137} (1986) 109. 


\bibitem{Teitelboim-87}
C.~Teitelboim and J.~Zanelli,
\newblock Class. Quant. Grav. {\bf 4}, L125 (1987).

\bibitem{Choquet-Bruhat-88}
Y.~Choquet-Bruhat,
\newblock J. Math. Phys. {\bf 29}, 1891 (1988).

\bibitem{Deruelle-03}
N.~Deruelle and J.~Madore, [arXiv: gr-qc/0305004].



\bibitem{GGGW} C. Garraffo, G. Giribet, E. Gravanis and S. Willison,
J. Math. Phys. \textbf{49} (2008) 042503, [arXiv:0711.2992].


\bibitem{Garraffo2} C. Garraffo and G. Giribet, Mod. Phys. Lett. {\bf A23} (2009) 1801, [arXiv:0805.3575]



\bibitem{Boulware:1985wk}  D.~G.~Boulware and S.~Deser,
Phys.\ Rev.\ Lett.\ \textbf{55}, 2656 (1985).

\bibitem{Wheeler} J. T. Wheeler, Nucl. Phys. {\bf B268}, 737 (1986);
Nucl. Phys. {\bf B273}, 732 (1986).


\bibitem{Cai:2001dz}  R.~G.~Cai,
Phys.\ Rev.\ D \textbf{65}, 084014 (2002), [arXiv:hep-th/0109133].



\bibitem{Charmousis}
  C.~Charmousis and J.~F.~Dufaux,
  Class.\ Quant.\ Grav.\  {\bf 19}, 4671 (2002),
  [arXiv:hep-th/0202107].


\bibitem{Zegers}
 R.~Zegers,
  J.\ Math.\ Phys.\  {\bf 46}, 072502 (2005),
  [arXiv:gr-qc/0505016].

\bibitem{DF} S. Deser and J. Franklin, Class. Quant. Grav. \textbf{22}, L103
(2005), [arXiv:gr-qc/0506014].


\bibitem{Dotti}
  G.~Dotti, J.~Oliva and R.~Troncoso,
  Phys.\ Rev.\  D {\bf 76}, 064038 (2007),
  [arXiv:0706.1830];  Phys.\ Rev.\  D {\bf 75}, 024002 (2007)
  [arXiv:hep-th/0607062].









\bibitem{Davis:2002gn}  S.~C.~Davis,
Phys.\ Rev.\ D \textbf{67}, 024030 (2003), [arXiv:hep-th/0208205].

\bibitem{Gravanis:2002wy}
E.~Gravanis and S.~Willison,
Phys.\ Lett.\ B \textbf{562}, 118 (2003), [arXiv:hep-th/0209076].

\bibitem{I} W. Israel, Nuovo Cim. \textbf{B44S10} (1966) 1 [Erratum, Nuovo
Cimento \textbf{B48} (1967) 463].



\bibitem{Gravanis:2007ei}
  E.~Gravanis and S.~Willison,
  Phys.\ Rev.\  D {\bf 75}, 084025 (2007), [arXiv:gr-qc/0701152].







\bibitem{Vacuum bubbles}
V. A. Berezin, V. A. Kuzmin and I. I. Tkachev, Phys. Lett. {120B},
(1983) 91; K. Maeda, Gen. Rel. Grav. {\bf 18}, (1986) 931;  H. Sato,
Prog. Theor. Phys. {\bf 76} (1986) 1250; S.~T.~Blau,
E.~I.~Guendelman and A.~Guth, Phys. Rev. {\bf D 35}, 1747 (1987);
A.~Aguirre and M.~C.~Johnson, Phys.\ Rev.\  D {\bf 72}, 103525
(2005) [arXiv:gr-qc/0508093];
  S.~V.~Chernov and V.~I.~Dokuchaev,
  [arXiv:0709.0616].





\end{thebibliography}
\end{document}